\documentclass[manuscript]{aastex}

\usepackage{graphics,graphicx}

\begin{document}

\title{DEM L241, a Supernova Remnant containing a High-Mass X-ray Binary}

\author{F. D. Seward}
\affil{Smithsonian Astrophysical Observatory, 60 Garden St.,
Cambridge MA 02138, USA}

\author{P. A. Charles}
\affil{School of Physics and Astronomy, University of Southampton,
  Highfield, Southampton, SO17 1BJ,UK and Department of Astronomy, 
University of Cape Town, Private Bag X3, Rondebosch 7701, Republic of 
South Africa}

\author{D. L. Foster}
\affil{South African Astronomical Observatory, P.O. Box 9, Observatory 
7935, Cape Town, South Africa and Vanderbilt University, Department of
Physics and Astronomy, Nashville, TN 37235}

\author{J. R. Dickel}
\author{P. S. Romero}
\affil{Department of Physics and Astronomy, University of New Mexico,
  1919 Lomas Blvd. NE, Albuquerque, NM 87131, USA}

\author{Z. I. Edwards}
\author{M. Perry}
\author{R. M. Williams}
\affil{Department of Earth and Space Sciences, Columbus State
  University,  Coca Cola Space Science Center, 701 Front Avenue,
  Columbus, GA 31901, USA}

\begin{abstract}
A Chandra observation of the Large Magellanic Cloud supernova remnant
DEM L241 reveals an interior unresolved source
which is probably an accretion-powered binary.  The optical
counterpart is an O5III(f) star making this a High-Mass
X-ray Binary (HMXB) with orbital period likely to be of order tens of days.
Emission from the remnant interior is thermal and spectral
information is used to derive density and mass of the hot material.  
Elongation of the remnant is unusual and possible causes of this are
discussed.  The precursor star probably had mass $> 25 M_{\odot}$
 
\end{abstract}

keywords: supernova remnants--Magellanic
Clouds--X-rays:binaries--X-rays:individual(DEM L241)

\section{Introduction}
In the Magellanic Clouds there are now 34 supernova remnants known to
emit X-rays.  Ten of these have interior pulsar-wind nebulae (PWNe) or compact
objects which also radiate in the X-ray band.  Although the sample is
small, there is great diversity.  Several manifestations of neutron stars are
represented and the present observation may be an example of yet
another.  

X-rays from the supernova remnant SNR
0535-67.5 in the H {\footnotesize II}
 region DEM L241 (Davis et al 1976) were first
detected in 1979 by Long et al (1981).   The supernova remnant was
first identified by Mathewson et al (1985) and mapped using the
optical [S {\footnotesize II}] emission as seen in 
Figure \ref{fig-opt-snr} which 
shows the remnant 
and the surrounding H {\footnotesize II} emission.
Since the initial Einstein observation there have been X-ray
detections by ROSAT (Williams et al. 1999), XMM (Bamba et al. 2006),
and now Chandra.  The XMM data provided the first detailed X-ray image and
showed elongated diffuse emission filling the area outlined by [SII]
filaments with a bright hard point source centered 
in the SE section.  Because the spectrum of this source was a power
law, Bamba et al (2006) identified it as an unresolved PWN.  The
Chandra observation was planned to resolve this object and to
distinguish the expected point-like pulsar from surrounding diffuse
emission.  The Chandra result indeed shows a clear point-like hard
source but there is no sign of a PWN close to or surrounding the point source. 


\section{The Chandra Observation} 

Chandra observed DEM L241 on 2011 February 7 and 8  for 46 ks with the
ACIS detector.  The observation was in two continuous parts of length
22 and 24 ks (OBSIDs 12675 and 13226) separated by an interval of 40
ks.  Figures \ref{fig-x-snr} and \ref{fig-x-contours}
show the result.  The supernova remnant has dimension $2.2^{\prime}
\times 5.3^{\prime}$ with diffuse emission filling the interior and
with no limb brightening from an outer shell.  The remnant appears to
be almost
divided into two parts.  The southern section, called the ``Head'' by Bamba et
al (2006) has dimension approximately $1.4^{\prime} \times 1.8^{\prime}$ with the
bright compact source somewhat off-center.  The elongated northern part,
called the ``Tail'' has dimension $2.2^{\prime} \times 3.7^{\prime}$.
The entire X-ray remnant fits within and follows the [S {\footnotesize II}] filaments
seen in Figure \ref{fig-opt-snr}.  The compact source, being close to
the brightest diffuse emission, appears to be associated with the
remnant but its characteristics are unusual, as will be discussed in
the next section.

\section{The Compact Source} 
\subsection{X-rays}

The compact X-ray source appears point-like and has a luminosity
$\sim$2$\times$10$^{35}$ erg s$^{-1}$ at the LMC distance of
50 kpc.  No extended PWN is
visible in the immediate vicinity and
the diffuse emission close to the point source is thermal.  As shown in Figure
\ref{fig-point}, the Chandra telescope PSF (Point Spread Function)
accounts very well for the appearance of the
source.  The source X-ray spectrum is a power law and it is appreciably
harder than that from the diffuse parts of the remnant.  

\begin{figure}
\center
\includegraphics[width=4in]{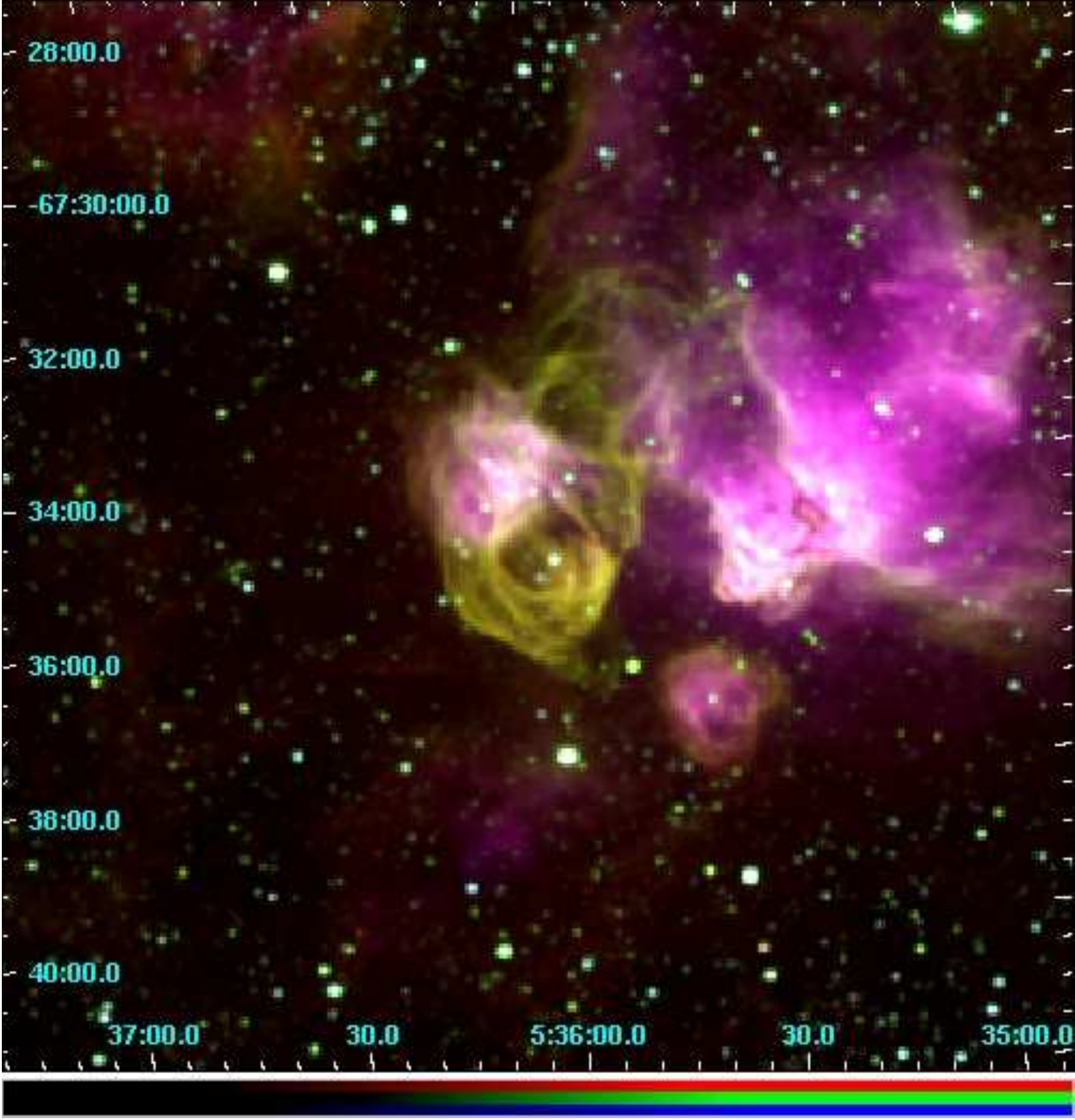}
\caption{\it The H {\footnotesize II} region DEM L241 showing H emission
  in red and [S {\footnotesize II}] emission in yellow.  The [S
    {\footnotesize II}] emission defines the supernova remnant and
  correlates well with the X-rays.  Figure from Smith et al 1999. \label{fig-opt-snr}}
\end{figure}

\begin{figure}
\center
\includegraphics[width=4in]{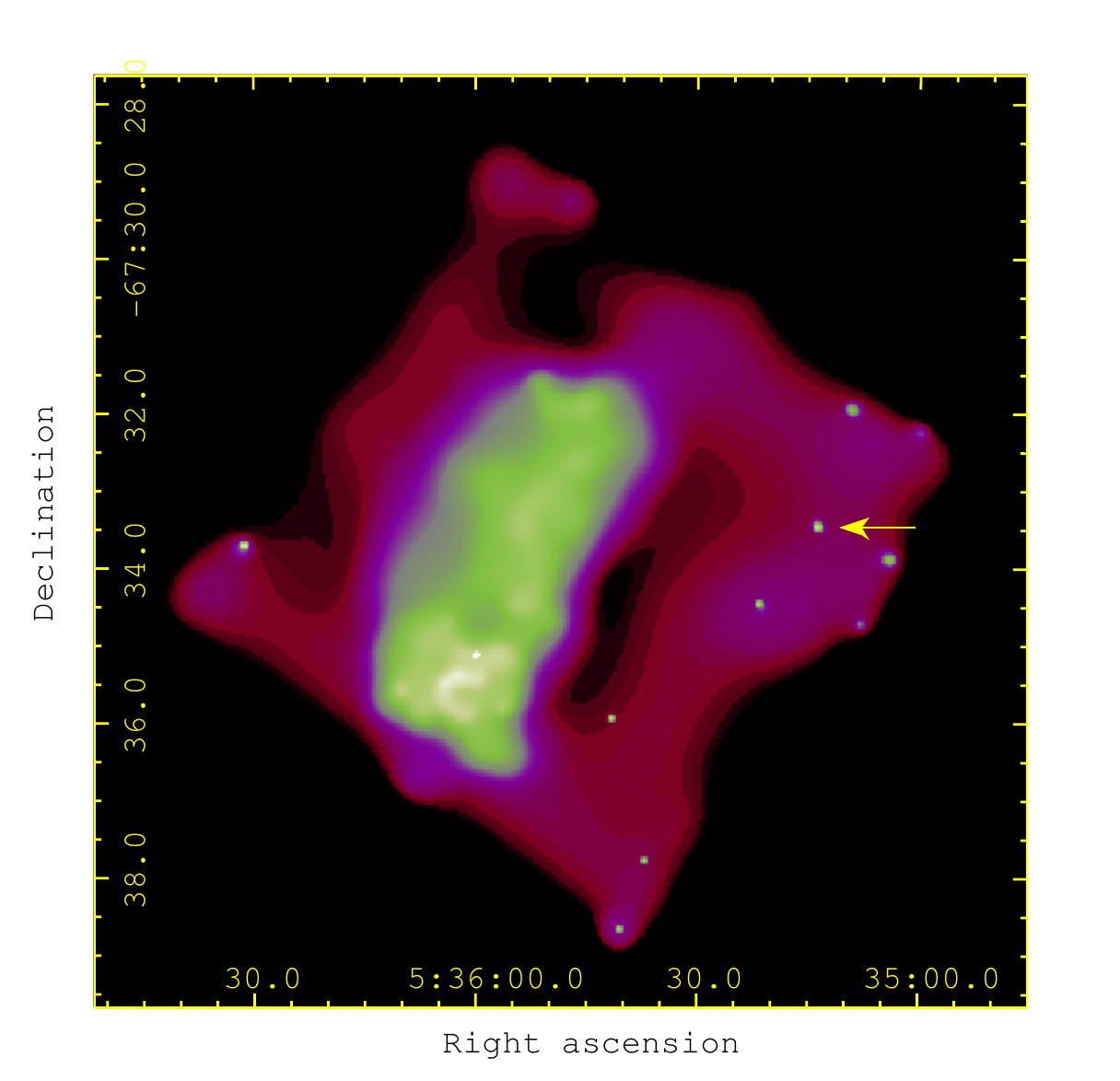}
\caption{\it  The field of the $8^{\prime} \times 8^{\prime}$ Chandra S3
chip.  This is an adaptive smoothing of X-ray data in the energy 
range 0.3 - 3 keV.
The color map shows increasing surface brightness going
from red to green to white.  The scale has been set so that the
supernova remnant in this figure is green. The
  point source in the SE part of the remnant is quite bright
  and is unresolved by Chandra.  The next Figure better
shows the true prominence of this source. Other unresolved sources in
  this field are foreground or background objects.  One of these, 
the star HD 269810, is indicated by an arrow and serves as a fiducial
reference. \label{fig-x-snr}}
\end{figure}

\begin{figure} 
\center
\includegraphics[height=4in]{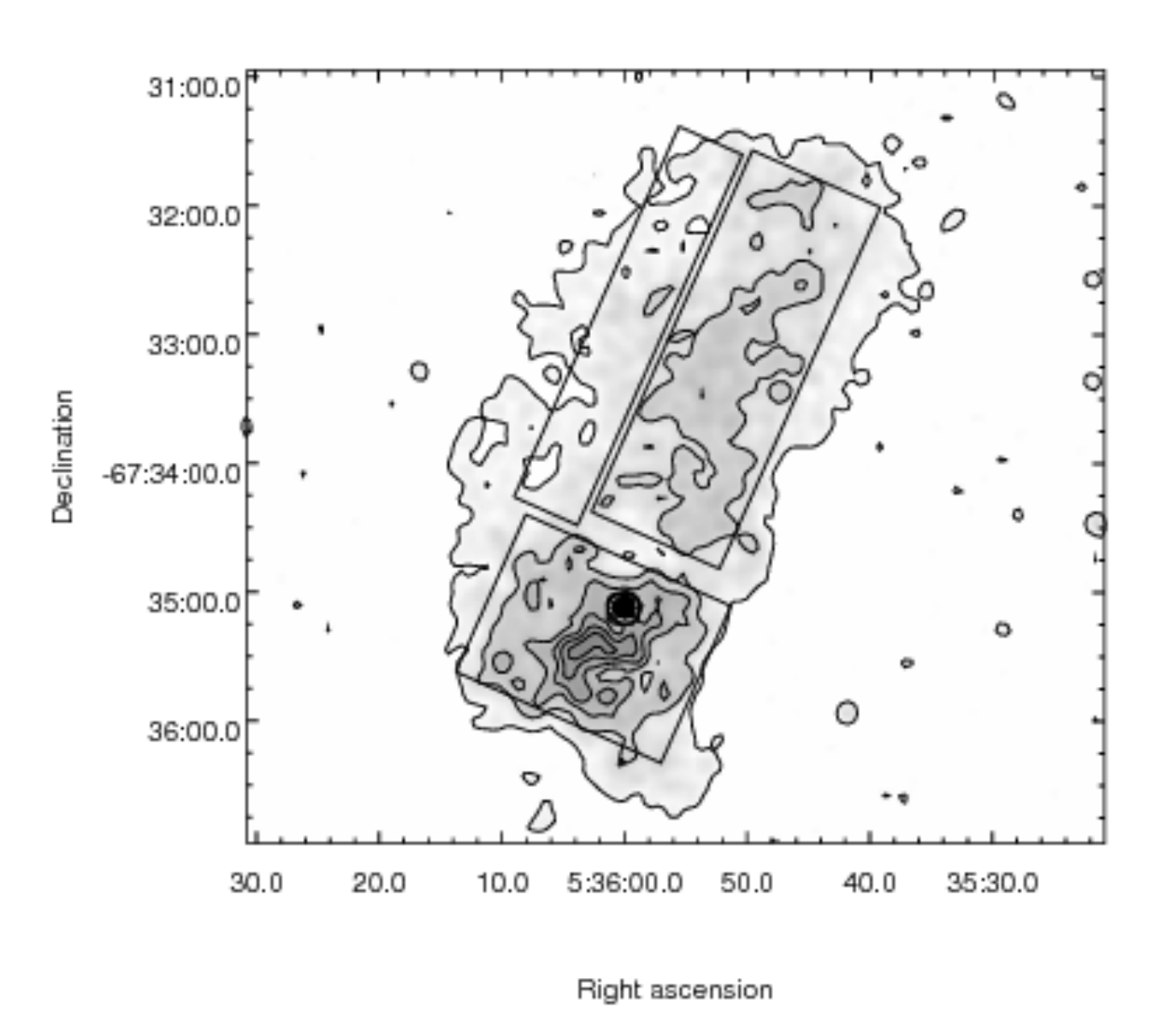}
\caption{\it Chandra observation of the remnant showing
contours of constant 
X-ray surface brightness overlaid on a smoothed image.  Smoothing is
Gaussian with $\sigma = 5^{\prime\prime}$.  Contours are linear with
separation 0.3 counts arcsecond$^{-2}$.  Note the diffuse structure just south of
the bright unresolved source.  The three boxes show spectral
extraction regions. \label{fig-x-contours}}
\end{figure}

\begin{figure}
\center
\includegraphics[width=4in]{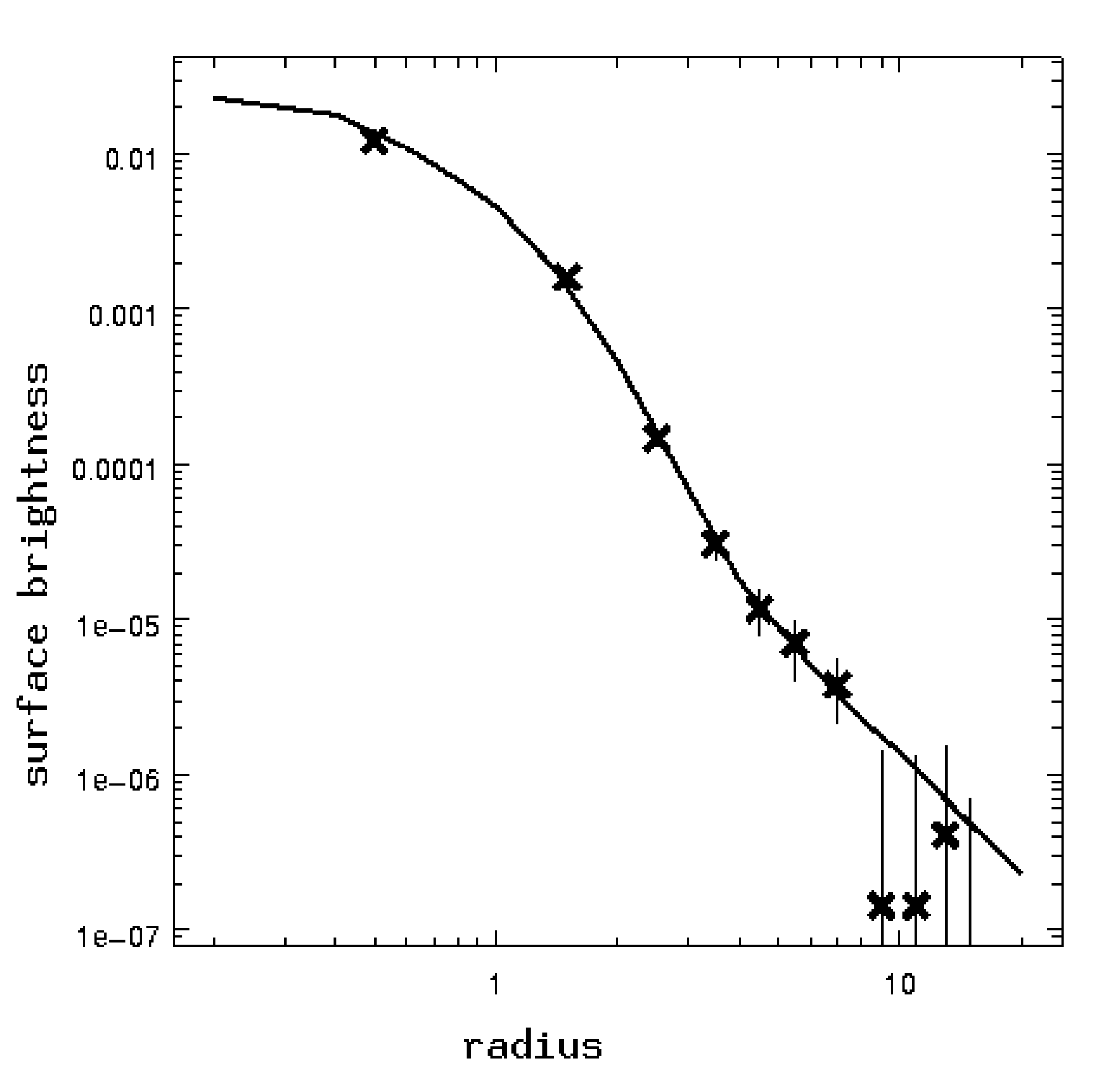}
\caption{\it Radial surface brightness of CXOU 053600.0-673507 compared
  with the Chandra telescope point spread function (PSF).  The diffuse
  SNR emission has been subtracted.  Data points are crosses
  with 1 $\sigma$ uncertainties due to counting-statistics.  The solid curve is
  the Chandra mirror PSF from Jarius 2002.  Units of radius are ACIS
  pixels (=0.492 arcsecond).  Units of surface brightness are counts
  s$^{-1}$ pixel$^{-1}$. \label{fig-point}}
\end{figure}

It is very unlikely that the object is a background AGN.  There are an 
average of 0.2 AGN deg$^{-2}$ at this flux level 
(Kim et al 2007) so the chance of finding one even within the $2^{\prime}
 \times 5^{\prime}$ remnant is only
$\approx 6 \times 10^{-4}$.  Furthermore the X-ray spectral index
 is -1.3 which is harder than the mean index of -1.9 
(with dispersion of 0.5) for radio quiet QSOs.  

The X-ray spectrum is also too hard to be that of a foreground normal 
star or even that of
 the optical counterpart O5III(f) star. This star itself is expected to be
 an X-ray source but with a soft spectrum and luminosity only 1\% of
 the flux we observe here (Chlebowski 1989). Colliding winds in a
 binary system could be more energetic but the luminosity of observed
 systems (e.g. W140 - Pollock et al 2005) is still an order of
 magnitude less than that of this object.   We conclude that the
 source is neither a foreground nor a background object and is probably
associated with the remnant. 

\begin{figure}[h]
\center
\includegraphics[width=4in]{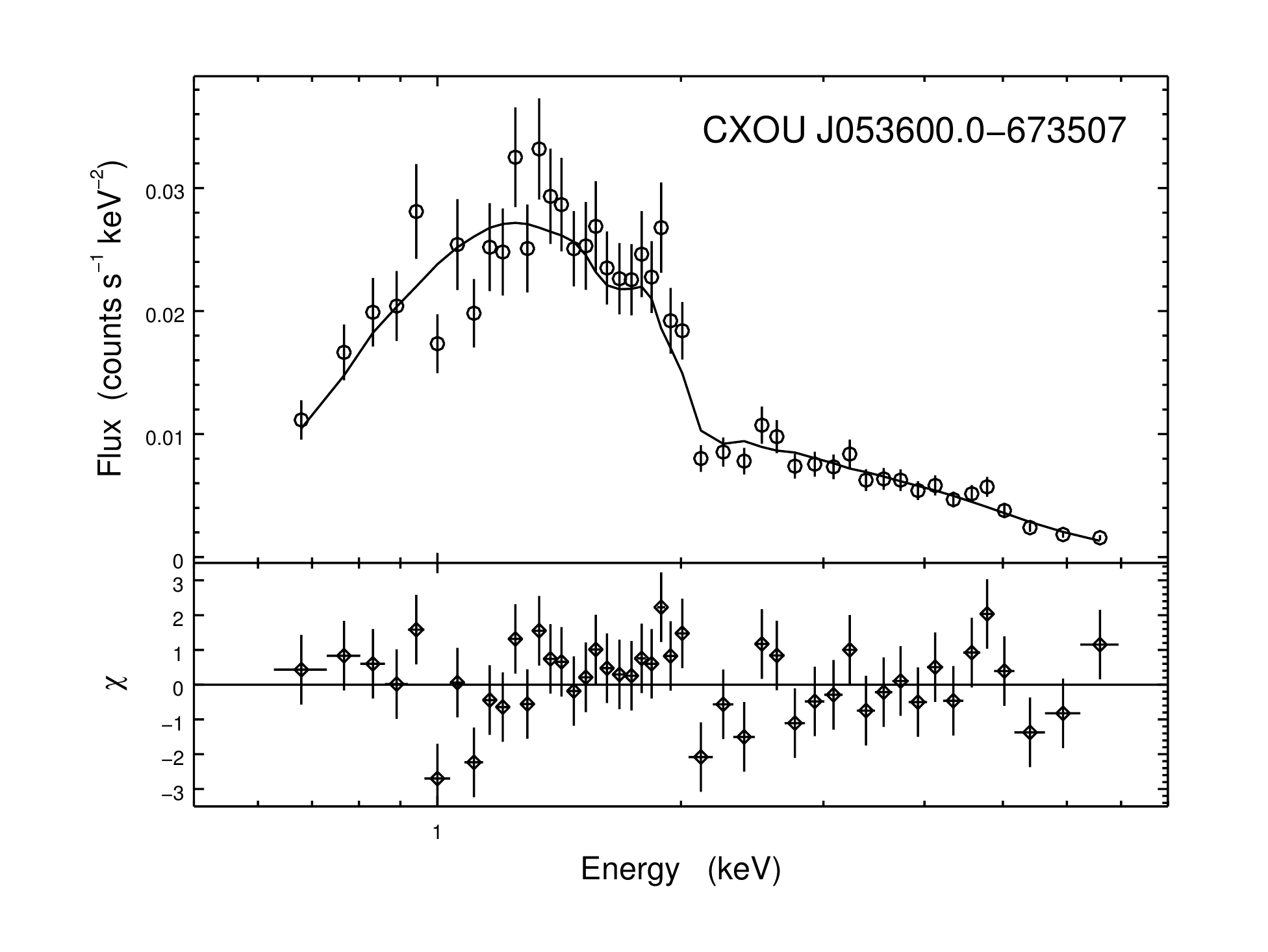}
\caption{\it X-ray spectrum of the compact source.  The solid line is the
  best-fit power law with pileup correction and the lower panel illustrates
  deviation of data points from this fit.  Uncertainties are $1
  \sigma$ in number of counts. \label{fig-compact-x-spec} }
\end{figure}

The Chandra X-ray spectrum is shown in Figure
\ref{fig-compact-x-spec}.  Data are from a circular region with radius
4$^{\prime\prime}$ centered on the source and background was taken from a
surrounding 5$^{\prime\prime}$ wide annulus.  The count rate was 0.057
s$^{-1}$ which, in the ACIS instrument which integrates for 3.2 s, leads
to a 3\% chance of two events being recorded as one.  This pileup is
not severe but enough to distort the spectrum.  We therefore incorporated
a pileup correction, {\it jdpileup}, in the Chandra {\it CIAO/SHERPA} spectral
analysis.  Results
are listed in Table 1.  The best fit was a power law with index
-1.28 $\pm 0.08$ and is shown in the figure.  Without the pileup
correction the best-fit index was -1.10 $\pm 0.06$.  (The uncertainties
given are $1 \sigma$.) 
The index measured with XMM,
free of pile up but with larger background subtraction, was
-1.57$ \pm 0.05$ (Bamba et al 2006).  The lack of agreement could be
due to an inadequate pileup correction, larger uncertainties than
quoted, or a variable source.  The Chandra and XMM-measured
luminosities are the same.  The luminosity range in Table 1 is due to
the observed variability.  We also fit a {\it xsdiskbb}
model to the data since this is appropriate for accretion powered
sources and there is reason to believe this object may be part of a
binary system.  The fit is reasonable but the absorption is too low.  
If this is an accretion powered system, a more elaborate model is needed.

\begin{table}[h]
\center
\title{Table 1.  ~~~ Compact source spectral fits}\\
\scriptsize
\begin{tabular}{llllll} \hline\hline
form&   energy range&   photon index&   ISM absorption&  reduced &   $L_x$ \\
       &(keV)                &      & $N_H$ ($10^{22}$)  &   $\chi^2$   &(erg s$^{-1}$)\\ \hline
power law&0.3-10   &$\Gamma = 1.28 \pm 0.08$&$0.19 \pm 0.034$& 1.26&$2.5-3.2 \times 10^{35}$\\
diskbb      & 0.3-10  &$kT_{in} = 2.43 \pm 0.23$   & $0.073 \pm 0.023$&1.38&$2.3 \times 10^{35}$\\ 
 \hline
\end{tabular}
\end{table}

\subsection{Location and variability}

The bright optical counterpart, a V=13.5 O5III(f) star, is easily 
visible within the Head of the remnant in Figure 1.  This had been noted as a possible
counterpart to the Einstein source CAL 60 (from the catalog of Long et
al, 1981) by Crampton et al (1985) who published a
finding chart and spectrum.   Since the
source CAL 60 includes the diffuse emission as well as the point
source we will refer to the point source as CXOU 053600.0-673507. Our
X-ray position is 2.2$^{\prime\prime}$ S of the XMM position and $<
0.6^{\prime\prime}$ from this O star.  X-ray positions are listed in
Table 2 with uncertainty the larger of that from counting 
statistics or the difference between positions from the two halves of
our observation. The systematic error associated with Chandra
positions is $0.6^{\prime\prime}$ at 90\% confidence. The star
HD2269810 is also in the field and our X-ray position for this star is
$0.5^{\prime\prime}$ from the optical and 2MASS IR survey position, so
the registration of the Chandra field is good and we adopt
$0.6^{\prime\prime}$ as the uncertainty of the X-ray position.

\begin{table}
\center
\title{Table 2.   ~~~Measured source positions}\\
\scriptsize
\begin{tabular}{llll} \hline\hline
source&waveband&RA (2000)&Dec (2000) \\ \hline 
CXOU J053600.0-673507&X-ray&05:36:00.01 $\pm$ 0:0:0.02 &-67:35:07.5 $\pm$ 0:0:0.2 \\
O star&optical&05 35 59.9 $\pm$ 0:0:0.08 & -67 35 06.3 $\pm$ 0:0:0.5\\
O star&IR&05:36:00.01&-67:35:07.6 \\ \hline
HD 269810&X-ray&05:35:13.82 $\pm$ 0:0:0.05&-67:33:28.0 $\pm$ 0:0:0.3\\
HD 269810&IR&05:35:13.89&-67:33:27.6 \\
HD 269810&TYCHO-2&05:35:13.90&-67:33:27.6 \\
\hline
\end{tabular}
\end{table}

The source is variable.
The count rate in the second part of the observation increased 25\% 
over that obtained in the first part.
No variability was seen on time scales shorter than $\sim 10^4$ s
although the observation was not sensitive to periods shorter than 6 s
 or to pulsed fractions $<$ 20\%.  This object cannot be an unresolved PWN as was
reasonably inferred from the XMM observation by Bamba et al.  

Since in $\approx 12$ hours the Chandra flux varied 25\% we searched
for variability in past
observations.  These were all  of limited sensitivity but could show
variability extremes.  Table 3 lists imaging observations which have
detected this remnant.   We inspected the archival Einstein and
ROSAT fields and, in all cases could distinguish the Head and Tail of
the remnant but could not separate the point source in the Head 
from the diffuse emission.   The
diffuse and compact parts of the remnant are only well separated in XMM
and Chandra observations. Indeed, the spatial resolution and low count rate
in Einstein and ROSAT observations make it impossible to do this if
the point source is no stronger than
in our observation.  In these past observations only 10-20\% of the counts in
the Head should be from the point source and total counts from the
Head range from $\sim$30 to  $\sim$200 with the higher rates from
detectors with the lower spatial resolution.  In all cases the 
counting rate was about
the same from the Head and Tail regions as it is in the Chandra
observation, so this is reassuring.

\begin{table}
\center
\title{Table 3.  ~~~ X-ray observations of DEM L241}\\
\scriptsize
\begin{tabular}{llccccl} \hline\hline
date&spacecraft&energy range&compact source&compact source& diffuse absorbed & reference \\
&&&0.5-5 keV abs. flux&$L_x$ 0.3-10 keV&0.5-5 keV Head flux & \\
&&(keV)&($10^{-13}$ erg cm$^{-2}$ s$^{-1}$)&($10^{35}$ erg
s$^{-1}$)&($10^{-13}$ erg  cm$^{-2}$ s$^{-1}$)& \\ \hline
1979 Apr 10&Einstein IPC&0.15-4&-&-&$4.1\pm 1.4$&Long et al 1981\\
1979 Nov 5&Einstein HRI&0.15-4.5&-&-&$5.4 \pm 1.4$&Mathewson et al 1985 \\
1991 Feb 14&ROSAT HRI&0.15-2.5&-&-&$6.0 \pm 1.6$&Schmidtke et al 1994\\
1993 Jul 24&ROSAT PSPC&0.15-2.5&-&-&$4.2 \pm 0.8$&\\
1995 Jul 9,17&ROSAT HRI&0.15-2.5&-&-&$5.1 \pm 1.0$&Schmidtke et al 1999\\
2004 Dec 29&XMM&0.5-10&&$2.32 \pm 0.14 $&&Bamba et al 2006\\
2011 Feb 7&Chandra&0.3-10&$3.71 \pm 0.10 $&$2.52 \pm 0.07 $&$4.42 \pm 0.12 $&this paper\\
2011 Feb 8&Chandra&0.3-10&$4.70 \pm 0.12 $&$3.19 \pm 0.08 $&$4.43 \pm 0.12 $&this paper\\
\hline
\end{tabular}
\end{table}

In order to compare observations of the diffuse source we have converted
count rates to flux levels with PIMMS, assuming a thermal spectrum
with $kT$ = 0.5 keV and $N_H$ = 2 $\times$ 10$^{21}$ cm$^{-2}$. To avoid
large absorption corrections at energies below 0.5 keV, we calculated
the flux in the range 0.5 to 5.0 keV and these results are summarized
in Table 3.  Uncertainties are counting statistics only.  The luminosity,
$L_X$, of the compact source is calculated for the Chandra range
0.3-10 keV.   The agreement of past observations is good, as 
indeed it should be for the diffuse emission.  Any variation of the point
source less than a factor of $\sim$3 is undetectable in the Einstein
and ROSAT data. The past observations of this remnant therefore show that
there have been no major outbursts at these times. 

We searched for regular pulsations in the Chandra data and found
nothing significant.  Since the ACIS
instrument integrates for 3.2 s, the search was only valid for periods
between 6.4 s and 6 h and the signal of 2500 counts limited the search to
pulsed fraction above $\approx$ 20\%.  Bamba et al. searched XMM and
ASCA data in the period range 0.15 - 500 s  at about this same
sensitivity with null result.  A short period and/or a $\approx$10\% pulse
fraction is quite possible.    

\subsection{Optical/NIR}

\begin{figure}[t]
\center
\includegraphics[width=4in]{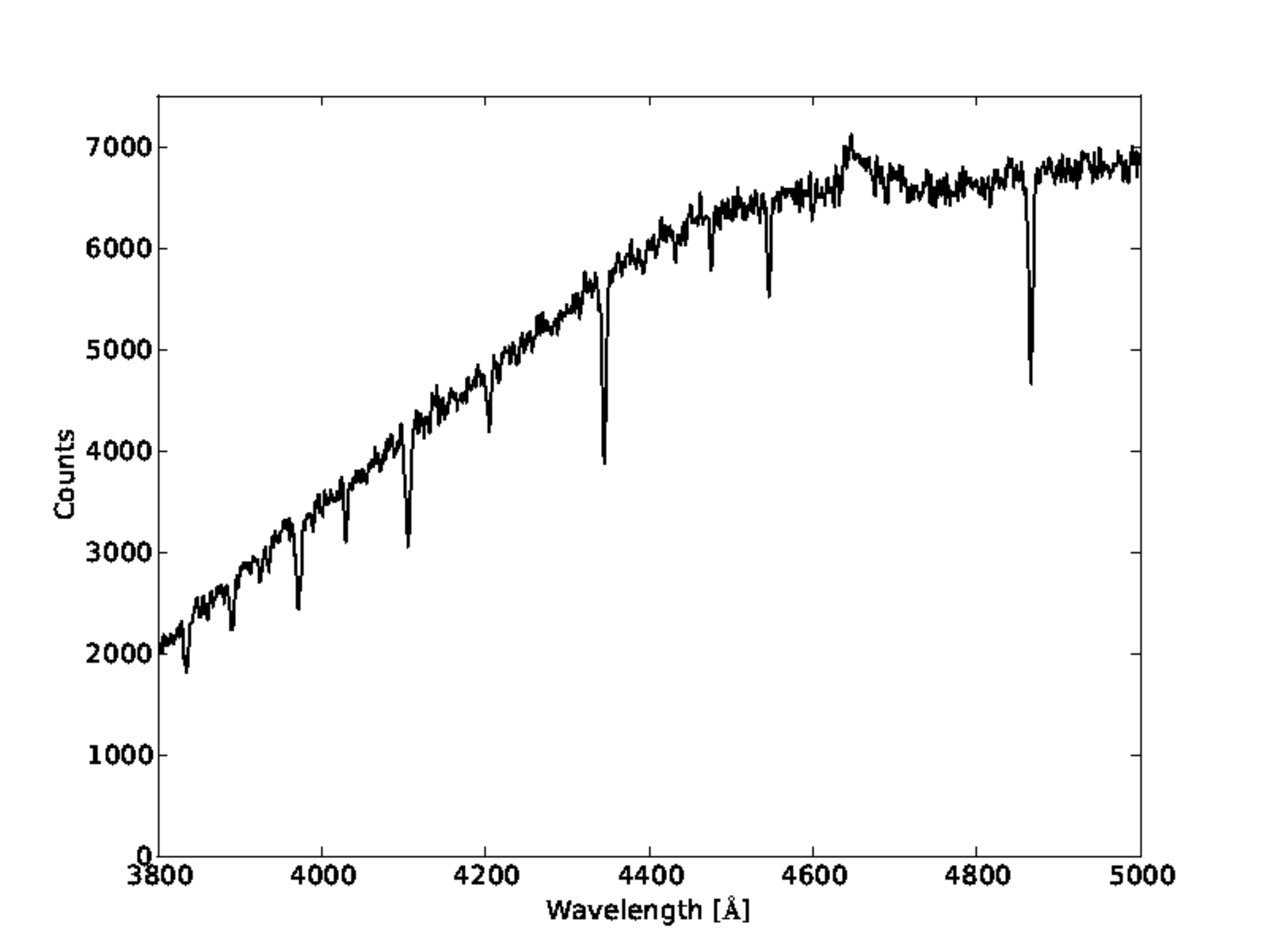}
\caption{\it Spectrum of the O5III(f) counterpart of CXOU 053600.0-673507 taken
  with the SAAO 1.9m telescope.  Absorption lines are from H and He.
  The emission feature is the Bowen blend, typical of very early O
  stars. \label{fig-opt-spec}}
\end{figure}

We have measured the optical spectrum of this star with the SAAO 1.9 m
telescope at Sutherland, South Africa in October 2011 and April 2012. 
The spectrum from 2011 Oct 24 is shown in Figure \ref{fig-opt-spec}. 
This spectrum is almost identical to
that of Crampton et al, indicating no long-term spectral
variability. The velocity we measure from the Balmer (mean of H-$\beta$ and
H-$\gamma$) absorption lines is 324$\pm$7 km s$^{-1}$ so it is clearly in
the LMC and an association with the remnant is possible. Further
spectra on 2012 April 12 and 15 yielded velocities of 301 and 296 km
s$^{-1}$ respectively (and with similar uncertainties). These values
are consistent with the Crampton et al (1985) mean value of 304 km
s$^{-1}$, as well as the variability by 34 km s$^{-1}$ that was noted
then. The small velocity change noted between our April spectra are
indicative of a long period system, likely tens of days or more. 

IR photometry in the J, H and K$_S$ wavebands of the optical
 counterpart was obtained 
during the month of November 2011 using the 1.4m IRSF telescope at
SAAO's Sutherland observing station. These showed no significant
variability to within 0.1 mag.

\section{The Diffuse Emission}

\subsection{Radio Observations}

\begin{figure}
\center
\includegraphics[width=4in]{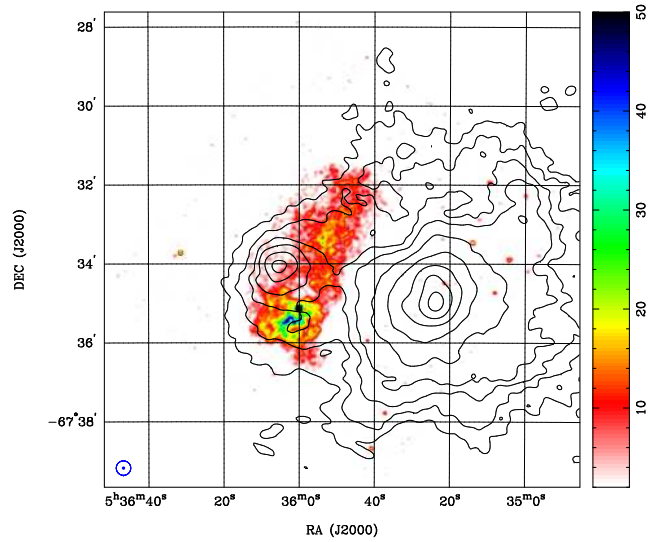}
\caption{\it Contours of constant 3.5 cm radio surface brightness overlaid
over the smoothed X-ray emission.  The X-ray field is that of Figure 
\ref{fig-x-snr} but with $5^{\prime\prime}$ Gaussian smoothing.  
Radio contours are drawn at 2, 3, 4, 5,
7, 10, 20, 50 \& 100 mJy/beam.  The dot in the lower left represents 
the 5-arcsec 
resolution of the smoothed X-ray image and the surrounding circle is the
20-arcsec beam width at 3.5 cm. The two radio maxima are from the bright 
H {\footnotesize II} regions seen
in Figure \ref{fig-opt-snr}. \label{fig-radio-x}}
\end{figure}

Moderate-resolution radio observations of this area were made during
surveys at  8.6- and 4.8-GHz (Dickel et al. 2005), at 1.4 GHz
(Hughes et al. 2007), and at 843 MHz (Mauch
et al. 2003).  The best resolution is at 8.6 GHz with a
half-power beam width of $20^{\prime\prime}$.  An overlay of the X-ray image on the 8.6 GHz radio contours is shown in Fig. \ref{fig-radio-x}.   The X-ray emission has been smoothed with a 5-arcsecond Gaussian.  The H {\footnotesize II} region is clearly delineated by the contours with its two  brightest parts N59A on the west and N59B on the east as designated by Henize (1956).  Because of the overlap with the bright H {\footnotesize II} region it is difficult to separate a possible contribution of the SNR to the radio emission.  The rms noise should give a 5 sigma limit of about 1 mJy/beam but the contamination could make the limit a few mJy/beam.  The fact that no radio emission appears to match with the X-ray emission is unusual.  Most SNRs with significant
X-ray emission are readily visible at radio wavelengths as well. Apparently the surrounding H II region complex masks the non-thermal emission from the SNR.

With the current resolution, any point radio source would be difficult
to distinguish from the somewhat irregular background, but the absence
of any feature at the position of the point source suggests that it would be less
than a few mJy.  For comparison, the X-ray only pulsar 0537-6910 in
N157B is less than .06 mJy at 22 cm (Crawford et al. 1998) and
PSR 0540-693 in the composite SNR of the same name is 0.4 mJy at 640
MHz. N206 is a shell-like radio SNR with a radio jet-like feature
with an X-ray point source at the tip, but no radio point source or
pulsar at $<$ .02 mJy (Williams et al. 2005).  Thus the radio limit on
the point X-ray source is not very significant or unexpected. 

\begin{figure}
\center
\includegraphics[width=4in]{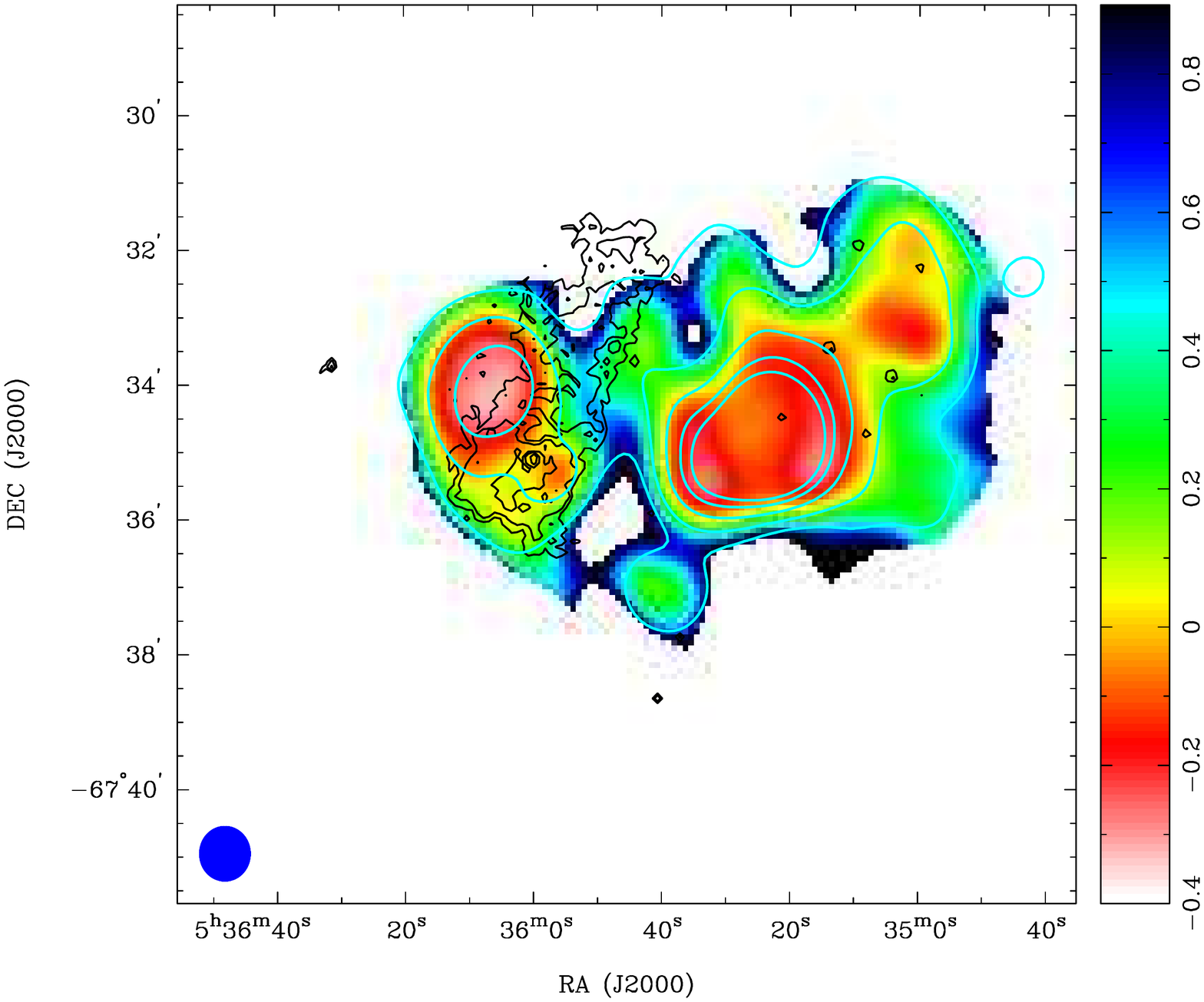}
\caption{\it  Color indicates the radio spectral index with red showing the
flat spectrum characteristic of thermal emission.  There is no
 indication of the steeper spectrum expected from the shell of 
a supernova remnant.  Smoothed contours of constant 21.7 cm radio 
surface brightness are overlaid and
drawn at 10, 20, 40, 70, \& 100 mJy/beam where the 48.5"x45" radio 
beam is indicated at the bottom left of the figure. Black contours 
show the X-ray surface brightness at levels of 0.25, 0.5, 1, 2.5, 
and 12 counts arcsecond$^{-2}$. \label{fig-radio-spec} }
\end{figure}

Another way to distinguish between the thermal emission from H {\footnotesize II}
regions and the non-thermal synchrotron emission from SNRs is the
radio spectrum. H {\footnotesize II} regions have a nearly flat
spectrum with a slope, or spectral index,
near 0 while PWN have a modest decrease in brightness as frequency
increases with an index of about -0.2 to -0.3.  Shell SNRs have
average spectral index slopes of -0.5 with young remnants being steeper
and old ones flatter.   Figure \ref{fig-radio-spec} shows a four-wavelength radio
spectral index map of the entire region.  Production of this image
required convolution of all the data to the 48.5 x 45.0 arcsec
resolution of the 36-cm image. Obviously, part of the X-ray Tail is
missing in the radio but other areas could contain some synchrotron
emission from a PWN or very old SNR in addition to the thermal
emission from the H {\footnotesize II} regions.  From the spectral index map, we
cannot distinguish the separation of various components.

\subsection{Optical Observations}

For comparison with optical wavelengths, we used images from the
Magellanic Cloud Emission Line Survey (MCELS; Smith 1999).  These
images were taken at the UM/CTIO Curtis Schmidt telescope at Cerro
Tololo Inter-American Observatory (CTIO).  The detector, a Tek 2048 x
2048 CCD with 24 micron pixels, gave a scale of 2.3 arcsec per pixel
and a resulting angular resolution of approximately 2.6 arcsec.  The
narrow band images were taken with filters centered on the [O III]
(5007 \AA, FWHM = 40 \AA), H-alpha (6563 \AA, FWHM = 30 \AA), and [S II] (6724
\AA, FWHM = 50 \AA) emission lines. The image shown in Figure
\ref{fig-opt-snr} has been reduced
using the IRAF software package for bias subtraction and flat-field
correction, and the astrometry was derived based on the HST Guide Star
and USNO-A catalogs.  This image, however, has not been
flux-calibrated or continuum-subtracted. 

\begin{figure}
\center
\includegraphics[width=4in]{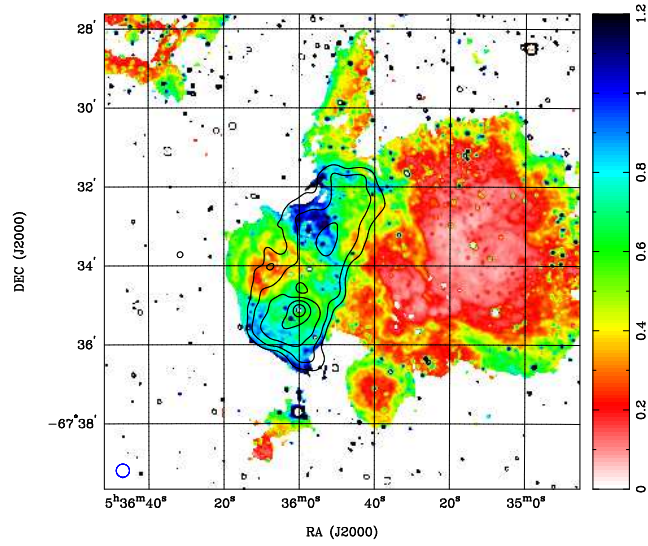}
\caption{\it Contours of constant X-ray surface brightness overlaid on the
ratio of [S {\footnotesize II}] to Balmer optical
emission.  X-ray data have been smoothed
with a $20^{\prime\prime}$ Gaussian and X-ray contours are drawn at
levels of 0.12, 0.25, 0.5, 1 \& 2 counts arcsecond$^{-2}$. \label{fig-hii-sii} }
\end{figure}

Because of the presence of the bright H {\footnotesize II} region, a good way to
depict the SNR is to show the [S {\footnotesize II}]/H$\alpha$ ratio. The [S II]
doublet at 671.7 and 673.1 nm is enhanced  by shocked material so any
region with an [S {\footnotesize II}]/H$\alpha$ ratio greater than about 0.5 is
considered a supernova remnant. In Fig. \ref{fig-hii-sii} we show the
ratio superimposed on contours of the X-ray emission.  There is
clearly a good correlation except on the eastern edge where the bright
H {\footnotesize II}-region component N59B overwhelms emission from the remnant. 

No X-rays are observed from the outer shock traced by the [S
{\footnotesize II}] filaments.  This implies that the temperature
of hot gas in the shock is $< 8 \times 10^5$ K.  The Sedov model then, can
be used to estimate an age limit.  Assuming $E_0 = 10^{51}$ ergs and
using the densities given in Table 5, we derive an age of $> 5-7 \times
10^4$ years.

\subsection{X-rays}

\begin{figure}
\center
\includegraphics[width=6in]{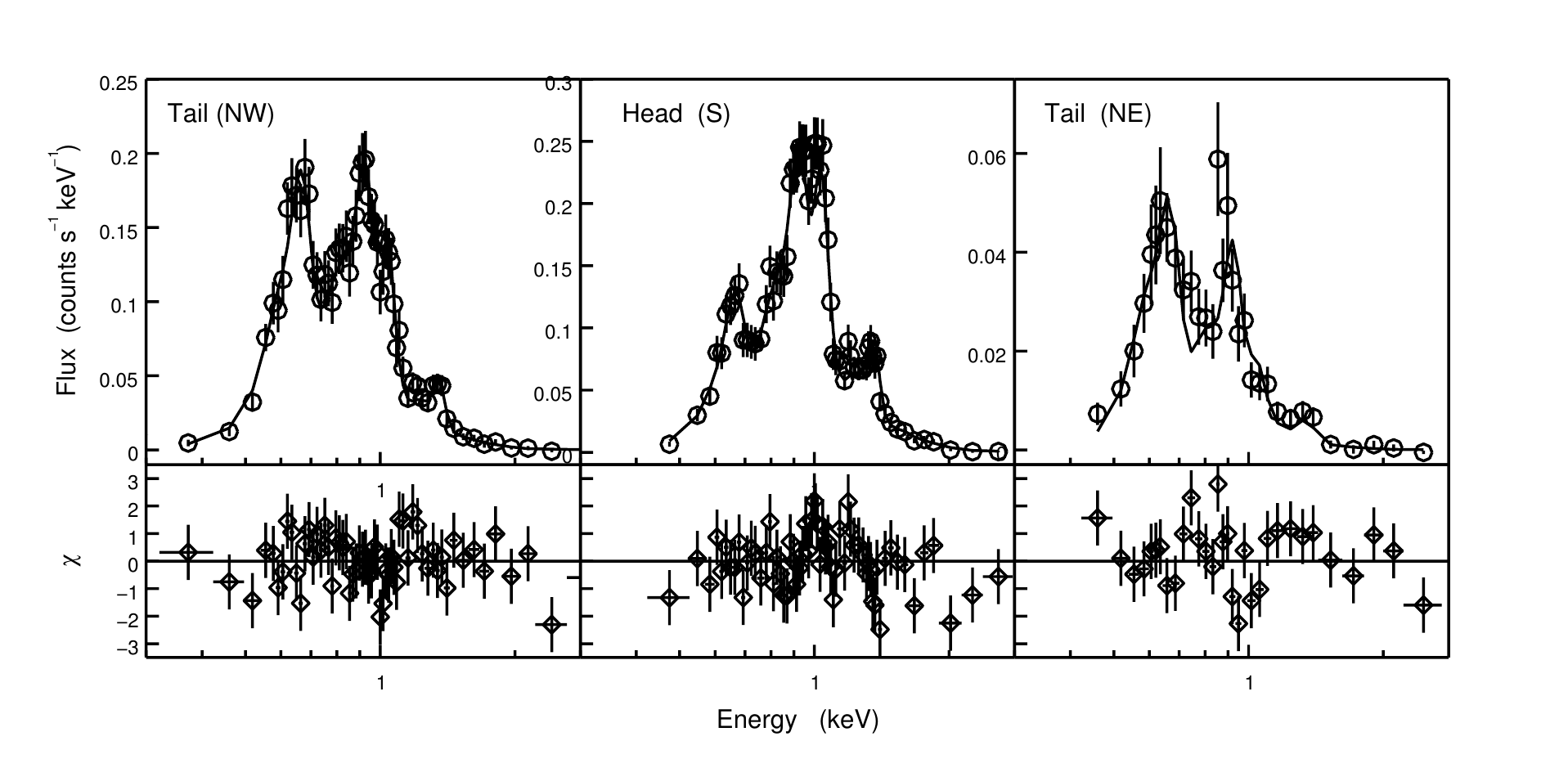}
\caption{\it X-ray spectrum of the three diffuse regions.  The lower panel 
illustrates
  deviation of data points from the best fit {\it xsvpshock model} shown by the
  solid lines.  A linear scale is used to show the large apparent
  differences between spectra from the three regions.  Only the 
absorption and timescale parameters vary from
  region to region.  Prominent emission lines are O at 0.67 keV, Ne and
  Fe at $\approx$ 1 kev, and Mg at 1.32 keV. \label{fig-diff-spec}}
\end{figure}

There are two major parts to this object.  The X-ray spectra of the
Head and the Tail are sufficiently similar to indicate that
they are parts of the same remnant rather than two separate objects
and we will so treat them.
Data were extracted from the 3 regions outlined in Figure
\ref{fig-x-contours} - the Head or S region, the bright NW strip  which
accounts for 80\% of the emission from the Tail and a NE strip where
there is fainter emission which might be indicative of a shock.  
After background subtraction the number of events in these regions
was: S - 7960, NW - 4640 and NE - 1020.  The
first two regions contain enough events to support an interesting
spectral analysis but the faint NE region yields less information.
Background was taken from large source-free regions northeast and 
southwest of the
remnant and was 35\% of the remnant signal in the S region, 85\% of the NW
signal and 200\% of the NE signal.
 The spectra of these regions are shown in Fig \ref{fig-diff-spec}. 
 The visually distinct spectra of the S and NW regions 
 can be explained almost entirely by
 differences in absorption of the emitted X-rays.  This
 can be seen by inspection of the softest emission and is greatest in
 the region between Head and Tail.  The visual gap in the X-ray
 image which separates the Head from the Tail is most prominent in the
 softest energy band and is caused by absorbing material. This
 is not surprising considering the complex surrounding environment.
The spectra of the bright structure just south of the O star and the
surrounding outer part of the Head were examined separately and were,
within uncertainties, the same.  The bright structure is therefore
plausibly explained as a density enhancement within the SNR or an
extension of the remnant structure along the line of sight.

We used the {\it xsvpshock} model, which accounts for non-equilibrium
ionization in a one-dimensional shock, to fit the NW region which shows
least absorption.  Metal abundances were fixed at 0.3 solar and
then
allowed to vary to fit the data.  It was necessary to vary only O, Ne,
Mg, and Fe.  We then fit the S and NE regions with the same
spectrum allowing only the ISM absorption and timescale parameters to
vary. Best-fit parameters are listed in Table 4.  The NW fit is
excellent, The S fit is good and the NE fit as good as any that can be
achieved with the limited number of events.  These fits are shown in
Figure \ref{fig-diff-spec}.  The difference between fits of the S and NW
regions is entirely due to different absorption of the O, Ne and Fe
emission lines.  The fit to the NE region requires a lesser value of 
 the ionization timescale ($\tau_{\mu}$).  Since the NE region
 comprises the outer part of the expanding remnant, this is expected.

The formal 1$\sigma$ uncertainties in the S and NW spectral parameters due to
counting statistics are $\approx$ 15\% for the 
absorption ($N_H$), $<10$\% for the temperature ($kT$), $\sim$ 40\% for
the abundances and $\sim$ 50\% for the ionization timescale
($\tau_{\mu}$).  The statistical uncertainties of the data 
points, however, allow acceptable fits over a large
range of parameters, e.g. the temperature can be varied by $\pm$ 50\%
and reasonable fits can be obtained with small changes in abundances
(although the Fe abundance is more sensitive than the lighter
elements).  Table 4 lists {\it CIAO/SHERPA} best-fit spectral
parameters for the {\it xsvpshock} and the equilibrium {\it xsvmekal} models.  

\begin{table}
\center
\title{Table 4.  ~~~ Diffuse source spectral fits}\\
\scriptsize
\begin{tabular}{l|lll|ll} \hline\hline
region & NW Tail & S Head & NE Tail & NW Tail & S Head \\
\hline
model &\multicolumn{3}{c|}{$xsvpshock$}&\multicolumn{2}{c}{$xsvmekal$}\\
\hline
reduced $\chi^2$ &0.79 & 1.05 & 1.38 & 0.91 & 1.08 \\ 
$N_H$ & 0.14 & 0.37 & 0.17 & 0.22 & 0.29 \\
\hline
kT & \multicolumn{3}{c|}{0.52} & 0.33 & 0.41 \\
O & \multicolumn{3}{c|}{1.9} & 1.2 & 2.9 \\
Ne & \multicolumn{3}{c|}{3.2} & 1.3 & 4.9 \\
Mg & \multicolumn{3}{c|}{1.7} & 1.0 & 2.3 \\
Fe & \multicolumn{3}{c|}{0.4} & 0.2 & 0.35 \\
\hline
$\tau_{\mu}$ & $7 \times 10^{11}$ & $7 \times 10^{11}$ & $2 \times
10^{11}$  &-&-\\
\hline
\end{tabular}
\end{table}

\section{Discussion}

\subsection{Nature of compact object}

The variability excludes an unresolved PWN such as that surrounding
the Vela Pulsar.  Furthermore the Vela PWN has $L_x = 2 \times 10^{33}$ ergs 
s$^{-1}$, considerably less than observed here.
The compact object is too bright to be a Cas A-type CCO 
(Central Compact Object with $L_x = 10^{33}-10^{34}$ ergs s$^{-1}$).
The luminosity is compatible with an AXP 
(Anomalous X-ray Pulsar with $L_x = 10^{35}-10^{36}$ ergs s$^{-1}$) 
but the spectrum is harder than that of most AXP (McGill Pulsar Group 2012).  
  The hard spectrum suggests an accretion-powered binary which,
comprising an O star and a neutron star, would be a High-Mass X-ray 
Binary (HMXB).  This is unexpected.  Only two other X-ray binaries
have been discovered within observable SNR -- SS 433, with a B or A
giant primary star,  within the
galactic remnant W50 and recently, a Be X-ray pulsar system, 
SXP 1062, within a faint Small-Magellanic-Cloud remnant
(H\'{e}nault-Brunet 2011).  

Although we have no evidence other than location that the
O star under discussion and the X-ray source form a binary system, 
we suspect that the object within DEM L241 is
the third HMXB to be found within a SNR.   The compact object, in this HMXB,
if a neutron star, would accrete
material from the O-star wind and changes in luminosity would be caused by
changes in the density of the wind.  Scaling from the HMXB Cen X-3, the
moderate luminosity and lack of observed outbursts imply (very
roughly) a separation of $\sim$ 1 AU and an orbital period of weeks.
At the moment this is speculation but the velocity of the O-star has
indeed been observed to vary.  No regular
pulsation in the X-rays have been found but
the observations were not sensitive to pulsed fractions below $\sim$ 20\%.

 The fact that the supernova precursor must have been more
massive than the O-star companion implies a precursor mass of 25, perhaps 40
$M_\odot$, thought to be the mass necessary for collapse to a black
hole.  It is therefore possible that the compact object, which we have assumed
to be a neutron star, could actually be a black hole.  
The X-ray luminosity, however, is unusual for a black
hole.  Known black hole binaries are either bright with $L_x \geq
10^{37}$ erg s$^{-1}$ or are transient systems in quiescence with $L_X
< 10^{33}$ erg s$^{-1}$ (Remillard and McClintock 2006). 

However, the HMXB in DEM L241 has a luminosity
($2 /times 10^{35}$ erg s$^{-1}$), intermediate between these extremes.
Interestingly, this is similar to the luminosity of SS433.  While
there is as yet no direct dynamical measurement of the compact object 
mass in SS433, all the indirect evidence strongly supports it being a 
$\sim$15 $M_\odot$ black hole accreting at extremely high mass
transfer rates from a highly evolved supergiant primary (see e.g. 
Blundell et al 2008).  The reason that it's X-ray luminosity is so low 
(${\sim}10^{35}$ erg s$^{-1}$) is that it is observed at high
inclination (it is eclipsing), and hence we only receive X-rays 
scattered from the surrounding material into our line of sight.  
i.e. it is an ``accretion disc corona'' system.
				   
As discussed earlier, there is marginal evidence that $L_x$ of
the HMXB in DEM L241 does not vary much, but we consider it unlikely 
that it too
would be a high inclination system.  Consequently, a black hole is certainly
possible but we have no strong indication that this might be the case.
A measure of the velocity variation of the O5III(f) companion would
provide a test of this hypothesis.  This remnant could become the
first example of a remnant known to be created by stellar collapse to
a black hole.

\subsection{Nature of the remnant}

Throughout this discussion it is assumed that the precursor star was
originally bound to the O5III(f) star.  After the SN, the compact remnant of
this remained bound to the O star to form the present HMXB.  The
explosion itself probably determined the size and temperature of the
remnant.  The binary environment of the precursor may have shaped the
remnant morphology.

The X-ray morphology of the SNR looks like that of a PWN with an
extended elliptical structure showing no obvious shell and a point
X-ray source near one end.  However, all diffuse emission, including that
close to the point source, is well fit with a thermal spectrum.  There
is no evidence for a power law component although a 10\% contribution
cannot be excluded.  The X-ray extent along the major axis of about 80 pc
makes the object one of the larger of the known SNR.  The SN may
have exploded between the two major H {\footnotesize II} regions and the expansion was
impeded in the east-west direction.  However, although
the SNR and the H {\footnotesize II} region complex do appear coincident in the sky
there is no indication in their structures that they have interacted
so they could be at different distances along the line of sight. 

The {\it vpshock} fits indicate an age of $\tau_{\mu}/n_e \sim 7
  \times 10^{11}/0.14-0.27$ s = $0.8-1.6 \times 10^5$ yr, about twice
  the lower limit estimated using the Sedov model.   This age 
is old for a radio SNR as the expanding shock will
have slowed sufficiently to reduce the shock acceleration of
relativistic particles that create the radio synchrotron emission.  Thus
the age can explain both the relatively flat radio spectral index
and the weak radio emission in general.

Since the plasma is close to
  equilibrium we also fit the X-ray spectra with the 
{\it xsvmekal} equilibrium model.
Table 4 gives the best fit result of this model to the S
and NW spectra.  In this case, the two spectra have been fit separately. 
The S region, the Head surrounding the compact object requires
somewhat higher temperature and abundances than the NW Tail region.
Absorption is still higher in the S than in the NW region. 

Although there is considerable uncertainty in the spectral fit
parameters and in the volumes used for the three regions, enhanced
abundances and derived densities indicate that we are seeing radiation
from supernova debris above that from the normal ISM in the Magellanic
Clouds.  Table 5 gives the plasma densities and masses calculated
from the {\it vpshock} results of Table 4.  The excess of O, Ne, and Mg
above that expected is 2.5, 0.8,and 0.15 $M_{\odot}$, indicating the
explosion of a massive star.  Woosley and Weaver (1995) have calculated the
masses of elements ejected from explosions of massive stars. This
amount of O can come from a precursor with mass $>25 M_{\odot}$ and
the Ne from a precursor with mass $>40 M_{\odot}$.  Since there is
variation in the different Woosley \& Weaver models and 
uncertainty in our measured abundances, we can only conclude that the
precursor was massive.  Since it evolved first, it should have been
more massive than the present O5III(f) star, so this is
expected.  Martins et al. (2005) give the mass of an isolated O5III(f)
star as about 40 $M_{\odot}$.  However, Rappaport \& Joss (1983) show
evidence that the companion stars in HMXBs are ``undermassive for their 
luminosity'', so a precursor mass estimate of $>25 M_{\odot}$ is more conservative.

\begin{table}[h]
\center
\title{Table 5.  ~~~ Properties of the remnant}\\
\scriptsize
\begin{tabular}{l|cccccc} \hline\hline
region & $F_x$ & $L_X$ & Volume & $n_e$ & Mass & $E_{th}$ \\
\hline
& ergs cm$^{-2}$ s$^{-1}$ & ergs s$^{-1}$ & cm$^3$& cm$^{-3}$ & $M_{\odot}$ &
ergs \\
\hline
S Head  & $4.17 \times 10^{-13}$ & $6.10 \times 10^{35}$ & $2.8 \times
10^{59}$ &0.27 & 83 & $1.9 \times 10^{50}$ \\
NW Tail & $3.86 \times 10^{-13}$ & $2.36 \times 10^{35}$ & $4.1 \times
10^{59}$ &0.14 & 61 & $1.1 \times 10^{50}$  \\
NE Tail & $0.83 \times 10^{-13}$ & $0.68 \times 10^{35}$ & $1.6 \times
10^{59}$ &0.12 & 21 & $0.4 \times 10^{50}$  \\
Total SNR &- & $9.14 \times 10^{35}$ & $8.5 \times 10^{59}$ &-& 165 & $3.4 \times 10^{50}$ \\
\hline
\end{tabular}
\end{table}

The remnant parameters derived here differ somewhat from those
obtained from the XMM observation by Bamba et al (2006).  Our X-ray
flux and luminosity is 30\% higher because we use a different energy band.
Our total mass is 30\% lower because we assume different volumes
for the emitting regions.  Bamba et al assume ellipsoidal volumes for
Head and Tail whereas we use cylindrical volumes with different
dimensions.  The Chandra best-fit spectra have higher temperature
and greater enrichment.  Remarkably, the total thermal energy content
is the same for the two analyses.

The elongated structure may be due to a ``blowout'' to a lower density
region in the north.  If so, this was a major event.  The Tail now has
twice the volume of the Head and contains 40\% of the thermal energy
of the hot gas.  On the other hand, perhaps the unusual  
elongation of the remnant is associated with
the unusual central source which is quite likely a HMXB. 
 The massive binary might
have produced a disk of material which channeled the ejecta into two
lobes.  Or, the precursor star itself could have been a Wolf-Rayet
star with strong wind which shaped a cavity now filled by the remnant.
Or, rotation of the precursor could have produced an
asymmetrical explosion.
 
\subsection{Comparison with W50/SS433}

There is an interesting parallel with the morphology of the
Galactic remnant W50 which is also large and elongated with dimension $95
\times 200$ pc (at a distance of 5.5 kpc, Lockman et al
2007). The binary SS 433 lies at the
center of W50 and produces opposing relativistic jets which have
inflated and elongated the remnant in the EW direction.  The optical
manifestation of these jets is spectacular (Margon et al 1980).   
SS 433 is an unresolved
X-ray source with luminosity $3 \times 10^{35}$ ergs s$^{-1}$, the
same as that of the DEM L241 source.   

DEM L241, with
dimension $32 \times 77$ pc, is also elongated but only about 
one third the size of W50. 
From the O star to the northern end of DEM L241 is 57 pc, about half the 
length of 120 pc for the longer arm of W50.   Perhaps the Tail of DEM L241
was inflated by a single northern jet and the structure south of the O
star is due to energy deposited by the oppositely-directed jet.  Since we see no other evidence for jets,
this probably happened in the past and is no longer operating.  The lesser size
of DEM L241 also indicates a weaker jet and/or a younger remnant.  

Despite intensive searches, no other examples of SS 433/W50 systems
have been found.  Here in the LMC is a system with similarities that
suggest further investigation might be worthwhile.
 
\subsection{Magellanic Cloud inventory}
  
As a matter of interest, we list here the compact objects now known to
be associated with Magellanic Cloud supernova remnants.  The Large and Small
Magellanic Clouds (LMC and SMC) contain $\sim 60$ remnants.
About 45 of these are known to emit X-rays
[Smith (web page,
2005), Williams et al. (2000), Inoue, Koyama, and Tanaka (1983),
Filipovic et al. (1998)].
Thirty-four have so far been observed by
Chandra and the results may be viewed in the online catalog at
http://hea-www.cfa.harvard.edu/ChandraSNR/.
Structure of the brighter remnants has been
well-resolved with moderate observing times and much detail is
visible.  One can see shock waves, interior structure,
clumpy material, central objects, and pulsar-wind nebulae.
Chandra data have been used to compile Table 6,
an inventory of X-ray emitting compact objects and PWN within
Magellanic Cloud SNRs. 
Diffuse thermal radiation from the shells or interiors has not been 
included.  Faint unresolved sources, for which no pulsations have
been detected, could contain a contribution from a small PWN so the
luminosities are listed under the header ``PSR \& PWN''.  
All luminosities have been corrected to the
0.5-10 keV range.  Ten of the 35 Magellanic Cloud SNRs observed by 
Chandra have PWN or compact objects.
The luminosity range is a factor of $10^4$ and there is great variety
of type; 1 SGR, 2 rotation-powered PSR/PWN, 1 CCO, 2 PWN with no
associated compact object, 2 XRB (including this observation), and 2
unclassified faint objects.

\begin{table}[h]
\center
\title{Table 6. Luminosity of Magellanic Cloud compact objects and PWNe} \\
\scriptsize
\begin{tabular}{lllllll}
\hline
\scriptsize
Remnant & PSR or CCO & period & $L_X$       & $L_X$  & $L_X$ & ref\\
        &            &        & PSR \& PWN & compact& pulsed& \\
\hline
        &            &  (s) & (erg s$^{-1}$) & (erg s$^{-1}$) & (erg s$^{-1}$)& \\
\hline
SNR 0540-693 & B0540-69 & 0.050 & $2.1 \times 10^{37}$ & $1.7 \times
10^{37}$ & $0.4 \times 10^{37}$ &(1) \\
N157B  & PSR J053747.3-691020 & 0.016 & $4.4 \times 10^{36}$ & $7
\times 10^{35}$ & $3 \times 10^{35} $&(2) \\
N49    & SGR J052600.8-660436 & 8.04 & $1.0 \times 10^{36}$&$1.0
\times 10^{36}$&$<10^{35}$? &(3) \\
DEM L241  &CXOU J053600.0-673507&-&-&$2 \times 10^{35}$&-&this paper \\
SNR 0453-685 & -&-& $6 \times 10^{34}$ &-&-&(4)\\
DEM L316B & - & - & $\approx 5 \times 10^{34}$ &-& &(5)\\
N23     &J050552.3-680141&-&-& $8 \times 10^{33}$ &-&(6)\\
SNR J0047.2-7308 & J004719.7-730823 &- & $\sim 3 \times 10^{33}$ & &- & (7)\\
N206   & - & -  &  $3 \times 10^{33}$& $ <1.5 \times 10^{33}$ &-&(8)\\
SNR J0127.7-7333&SXP 1062&1062&-&$7 \times 10^{35}$&$1.4 \times 10^{35}$&(9) \\
\hline
\multicolumn{7}{c}{(1) Kaaret et al 2001 (2) Wang et al 2001 (3) Kaplan 2002 (4) Gaensler et al 2003 (5) Williams \& Chu 2005} \\
\multicolumn{7}{c}{ (6) Hughes et al 2006 (7) Seward et al 2012 (8) Williams et al 2005 (9) H\'{e}nault-Brunet 2011} \\
\end{tabular}
\end{table}

\section{Conclusions}

This is an interesting supernova remnant, larger and more elongated than
most.  The absence of X-ray emission from the outer shock, the X-ray
spectra, and the size all indicate an older remnant.  Because of its
age and location close to a strong H {\footnotesize II} region, radio
emission is weak and undetectable.  It contains a compact object with an O-star
 optical counterpart which is unusual for a supernova remnant.
  No non-thermal X-rays, as might originate in a
 PWN, are detected from the immediate vicinity of this source and
 the source luminosity and spectrum are consistent with that expected
 from a HMXB.  The velocity of the O-star was observed to vary and a
 long period is suspected.

The diffuse X-ray spectrum from the remnant interior 
 is enriched in O, Ne, and Mg.  This enrichment and the more-slowly
 evolving O-star companion imply that the supernova precursor star had a
mass of $>25 M_{\odot}$.   We note that the elongated envelope 
and the structure close to the compact source might have been formed
by an SS 433-type pair of jets but there is no indication of these in
the optical spectrum. 

Support for this work was provided by the National Aeronautics and
Space Administration through Chandra Award Number GO1-12094 issued by
the Chandra X-ray Observatory Center, which is operated by the
Smithsonian Astrophysical Observatory for and on behalf of the
National Aeronautics Space Administration under contract NAS8-03060.
D.L.F. acknowledges support from NASA through the Harriett G. Jenkins 
Pre-doctoral Fellowship Program, and from the Vanderbilt-University 
of Cape Town Partnership.
We thank Paul Green for an interesting discussion and information 
about background quasars.  Sean Points supplied calibrated and aligned
MCELS images.
B. Furnish, J. Hood, C. McCarty, and T. Williams at Columbus State
University helped with the X-ray spectral analysis.

\section{References}

Bamba, A., Ueno, M., Nakajima, H., Mori, K., \& Koyama, K. 2006,  A\&A 450, 585 

Blundell, K.M., Bowler, M.G. \& Schmidtobreick, L. 2008, ApJ 678L, 47

Chlebowski, T. 1989, ApJ 342, 1091

Crampton, D., Cowley, A.P.,Thompson, I.B. \& Hutchings, J.B. 1985, AJ 90, 43

Crawford et al. 1998, Mem.de Soc. Astron. Italy. 69, 95

Davis, R.D., Elliot, K.H. \& Meaburn, J. 1976, MmRAS 81, 89 

Dickel, J. R., Mcintyre, V, Gruendl,,R., \& Milne, D. K. 2005, AJ 129, 790

Gaensler, B. M.; Hendrick, S. P., Reynolds, S. P., Borkowski, K. J. 2003, ApJ 594, L111 

H\'{e}nault-Brunet, V. et al 2011, MNRAS 420, L13

Henize, K. 1956, ApJS 2, 315

Hughes, A., Staveley-Smith, L., Kim, S., Wolleben, M., \&  Filipovi\'{c}, M. 2007, MNRAS, 382, 543.

Hughes, J.P., et al. 2006, ApJ 645, L117

Jarius, D. 2002, Comparison of on-axis Chandra Observations of AR Lac
to SAOSAC Simulations.  CXO internal memorandum.

Kaaret, P, et al., 2001, ApJ 546, 1159

Kaplan, D.L., 2002, MmSAI 73, 496

Kim, Minsun, et al. 2007, ApJS 659, 29

Lockman, F.J., Blundell, K.M. \& Goss, W.M. 2007, MN 381, 881

Long, K.S., Helfand, D.J., \& Grabelsky, D.A. 1981, ApJ 248, 925 
	
Margon, B., Grandi, S. A. \& Downes, R. A. 1980, ApJ 241, 306

Martins, F., Shaerer, D., \& Hillier, D.J. 2005, A\&A 436, 1049

Mathewson, D.S., Ford, V.L., Tuohy, I.R., Mills, B.Y., Turtle, A.J. \& Helfand, D. J. 1985, ApJS 58, 197  

Mauch, T., Murphy, T., Buttery, H. J., Curran, J., Hunstead, R. W.,
Piestrzynski, B., Robertson, J. G., Sadler, E. M, 2003, MNRAS 342,
117

McGill Pulsar Group 2012, McGill SGR/AXP Online Catalog, http://www.physics.mcgill.ca/~pulsar/magnetar/main.html

Pollock, A., Corcoran, M., Stevens, I. \& Williams P. 2005, ApJ 629, 482

Rappaport S. A. \& Joss P. C. 1983, in: W. H. G. Lewin \& 
E. P. J. van den Heuvel (eds.)
Accretion-driven X-ray Sources, (Cambridge: Cambridge University
Press), 33

Remillard, R. \& Mcclintock, J. 2006, ARA\&A 44, 49

Schmidtke, P. C., Cowley, A. P., Frattare, L. M., McGrath, T. K., Hutchings, J. B. \& Crampton, D. 1994, PASP 106, 843

Schmidtke, P. C.; Cowley, A. P.; Crane, J. D.; Taylor, V. A.; McGrath, T. K.; Hutchings, J. B.; Crampton, D. 1999, AJ 117, 927

Seward et al 2012, Chandra Supernova Remnant Catalog, http://hea-www.cfa.harvard.edu/ChandraSNR/

Smith, R.C. and the MCELS team 1999, IAU Symposium 190, 28

Wang, Q. D., Gotthelf, E. V., Chu, Y.-H., Dickel, J. R.  2001, ApJ 559, 275 

Woosley, S.E. \& Weaver, T.A. 1995, ApJS 101, 181

Williams, R.M. \& Chu, Y.-H. 2005 ApJ 635, 1077

Williams, R. M., Chu, Y.-.H., Dickel, J. R., Gruendl, R. A., Seward,
F. D., Guerrero, M. A., \& Hobbs, G. 2005, ApJ., 628, 704

\end{document}